# Electron-Phonon Coupling Mediated by Fröhlich Interaction in $Rb_2SnBr_6$ Perovskite

C. C. S. Soares [1], J. S. Rodríguez-Hernández [1], Bruno P. Silva [2], Mayra A. P. Gómez [1], V. S. Neto [1], A. P. Ayala [1], and C.W.A. Paschoal [1,*].

[1] Departamento de Física, Universidade Federal do Ceará, Campus do Pici, 65455-900, Fortaleza - CE, Brazil.

[2] Instituto Federal de Educação, Ciência e Tecnologia do Ceará, Campus Camocim, 62400-000, Camocim - CE, Brazil.

[*] Corresponding author. E-mail address: paschoal.william@fisica.ufc.br

## ABSTRACT

Due to their well-suited optoelectronic properties, metal halide perovskites are emerging semiconductor materials with potential applications in solar cells, detectors, and light-emitting diodes. Beyond the traditional 3D perovskites, low-dimensional counterparts have more attractive effects such as excitonic emissions and quantum confinements that are enhanced by the reduced dimensionality, which involve the Electron-Phonon Coupling (EPC). Such phenomenon, which comprehends the interaction between charge carriers and lattice vibrations, usually strongly impacts the photoluminescence (PL) response in low-dimensional frameworks. In this paper, we investigated the intrinsic EPC onto low-temperature PL of the zero-dimensional (0D) $Rb_2SnBr_6$ perovskite. Temperature-dependent PL measurements, complemented by various characterization techniques and theoretical calculations, revealed broadband emission with a significant Stokes shift attributed to self-trapped excitons (STEs). The Fröhlich mechanism, mediated by interactions between excitonic charge carriers and longitudinal optical (LO) phonons, primarily accounts for the emission broadening through phonon-assisted radiative recombination. The EPC strength was evaluated through the Huang-Rhys factor $S = 34$, confirming strong correlations between electronic and vibrational properties and supporting the STE emission assumption. The possible mechanism of STE formation was evaluated by the Fröhlich parameter $\alpha$ of 2.78 for electrons and 4.41 for holes, which points out a major contribution of the hole-polaron quasi-particle on exciton trapping. Our findings give insights regarding the influence of EPC in 0D perovskites and STE formation, which leads to the assessment of $Rb_2SnBr_6$ for light-harvesting applications.

# I. INTRODUCTION

For the past few years, Metal Halide Perovskites (MHP) have been extensively investigated due to their remarkable optoelectronic properties as tunable bandgap energy, intense light absorption, and efficient charge transport, which are convenient for photovoltaic applications such as solar cells, detectors, and light-emitting diodes (LEDs) devices [1–3]. Beyond traditional three-dimensional (3D) perovskites, low-dimensional variants comprising 2D, 1D, and 0D structures have attracted attention for their distinctive superior stability, excitonic behavior, and quantum confinement effects, which are one of the consequences of the reduced dielectric screening emerged from lowering dimensionality [4–6], resulting in stronger electron-hole interactions and larger exciton binding energies that can remain stable even at room temperature, leading to modified charge transport and enhanced radiative decay rates [7–9].

One of the most intriguing factors that influence the optoelectronic behavior of MHP is the Electron-Phonon Coupling (EPC), which underlies the interaction between charge carriers and lattice vibrations and has a compelling role in carrier mobility and phonon-assisted radiative recombination [10–16]. EPC is enhanced in low-dimensional MHP, often triggering carrier localization through a quasi-particle called polaron formed by the carrier and its own polarization field that can be thought of as an electron (or hole) moving slowly by dressing itself with a phonon cloud [17–20], and becoming localized. Although excitons are neutral quasi-particles, carrier localization can also restrain their movements along the lattice and create self-trapped excitons (STE) [21–23]. The electron-phonon interactions in polaron formation of polar perovskites can be described by theoretical models like the Fröhlich interaction [10,24–26], which accounts for the coupling between charge carriers and longitudinal optical (LO) phonons where the Fröhlich parameter $\alpha$ [27], dictates the interaction strength between carries with LO vibrations. In the case of excitonic emissions, the coupling strength can also be quantified by the Huang-Rhys factor $S$ [28,29], which can be extracted from temperature-dependent photoluminescence (PL) spectra by analyzing the phonon contributions to PL linewidth broadening [30,31]. Experimentally, temperature-dependent PL, Raman, and IR spectroscopies have been widely used to investigate EPC [32–34], providing essential information on phonon modes and their influence on the optical response.

Vacancy-ordered 0D perovskites depicted by the chemical formula $A_2BX_6$ offer a unique crystal structure feasible for investigating the EPC in low-dimensional MHP [35,36]. They consist of an isolated $[BX_6]$ octahedra arrangement among a framework composed of the A-cations. The isolated configuration provides a more favorable environment for charge carriers to interact with lattice vibrations since the octahedra are not as constrained together as their high-dimensional counterparts, amplifying the effects concerning the charge carriers-vibrations interplay and quantum confinement [37,38], which usually results in a broadband emission with large Stokes-shift from STE resulting in high PL quantum yields [39]. Furthermore, the optoelectronic phenomena concerning the 0D perovskites are intimately related to the isolated octahedra units where the surroundings A-cations not only have minimum effects on the electronic band structure of the octahedra (which maintains their photophysical properties) but also increase considerably these perovskites stability upon air exposure protecting the metal halide octahedral anions [35]. All these features have shown the potential applications of 0D perovskites in LED lighting, sensors, and X-ray scintillation [40,41].

Understanding the mechanisms involved in the EPC of 0D perovskites is essential for their optoelectronic applications, which creates opportunities for a range of possibilities, from the fundamentals to device engineering. Thus, in this paper, we investigated the EPC and its influence on the PL emission of $Rb_2SnBr_6$ vacancy-ordered 0D perovskite. We discuss the EPC strength, its main mechanism, and its major role on the PL origin and linewidth broadening with the aid of several characterization techniques and theoretical calculations. Our results elucidate strong EPC via Fröhlich interactions with LO phonons undergoing low-temperature STE emission of $Rb_2SnBr_6$ characteristic of polar inorganic perovskites.

## II. EXPERIMENTAL AND COMPUTATIONAL METHODS

**Synthesis Method:** Light-yellow $Rb_2SnBr_6$ single crystals were grown through the slow evaporation method: 48 mg (0.18 mmol) of $Rb_2SO_4$ (289280: Sigma-Aldrich) and 50 mg (0.18 mmol) of $SnBr_2$ (309257: Sigma-Aldrich) were dissolved using hydrobromic acid (47 wt% in $H_2O$) (5 mL) and distilled water (5 mL). The solution was heated at 80 °C for 1 hour under rigorous stirring and then cooled slowly to room temperature in a beaker sealed with paraffin film. Clear light-yellow microcrystals grown at the bottom of the beaker were collected and cleaned with ethyl ether and dried overnight at 60 °C in a vacuum furnace.

**Single Crystal X-Ray Diffraction:** Crystal structure determination was conducted using single crystal X-ray diffraction (SCXRD) performed on a Bruker D8 Venture X-ray diffractometer equipped with a Photon II Kappa detector and Mo K-alpha radiation ($\lambda$ = 0.71073 Å) microfocus source. A crystal sample was mounted on a MiTeGen MicroMount using immersion oil. The APEX 4 software was used for unit cell determination and data collection. The data reduction and global cell refinement were made using the Bruker SAINT$^+$ software package. The absorption correction was performed with SADABS [42,43]. The structure was solved by intrinsic phasing using SHELXT and refined by least squares on SHELXL under the Olex2 [44–46] graphical interface. The crystallographic artwork representations were prepared using VESTA software [47]. The determined crystal structure belongs to the $Fm\bar{3}m$ space group with $a = 10.66180(10)$ Å as cell parameter, consistent with Ketelaar *et al.* [48]. The $Rb_2SnBr_6$ crystals exhibited thicknesses ranging from 150 µm to 250 µm.

**Infrared Spectroscopy:** A Bruker Vertex 70 V Fourier-transform spectrometer acquired the infrared reflectance spectrum. Data in the far-infrared region were collected using a mercury (Hg) lamp as the light source, while signal detection was carried out using DLaTGS pyroelectric detectors. The dataset was obtained through 256 scans and a spectral resolution of 2 $cm^{-1}$.

**Diffuse UV-vis Reflectance Spectroscopy:** Room-temperature absorption spectra of grounded crystals were acquired using Diffuse Reflectance Spectroscopy (DRS) with a Shimadzu UV-2600 spectrophotometer equipped with an ISR-2600 Plus integrating sphere, covering the 200–800 nm wavelength range. The reflectance data were converted into absorption spectra via the Kubelka-Munk function [49].

**Photoluminescence (PL) and Raman Spectroscopies at Low-Temperatures:** PL spectra were collected using a T64000 Jobin–Yvon spectrometer equipped with an Olympus microscope and an $LN_2$-cooled CCD to detect the emitted radiation of the sample in a single mode. The temperature-dependent PL spectra were excited with an external lamp (405 nm) using a long working distance plan-achromatic objective of 20x and keeping the sample in a vacuum inside a He-compressed closed-cycle cryostat. A Lakeshore 330 controller controlled the temperature with the precision of ±0.1 K. Excitation intensity-dependent PL was recorded at 10 K using an Argon ion laser operating at a wavelength of 457 nm, digitally tuning the laser power and measuring its real value at laser focus with a Coherent LaserCheck Power Meter, keeping low intensities to avoid laser damage. The same temperature-dependent experimental setup was utilized to perform low-temperature Raman measurements of the material with a 457 nm laser line from Argon ion laser. Each Raman and PL spectrum was deconvoluted for data analysis based on a sum of Voight functions for Raman peaks and Gaussian functions for PL peaks using Fityk software [50].

**Density Functional Theory (DFT) Calculations:** First-principles calculations were performed using the Cambridge Serial Total Energy Package (CASTEP) code [51], within the framework of Density Functional Theory (DFT) [52,53]. The Generalized Gradient Approximation (GGA) functional described by Perdew−Burke−Ernzerhof (PBE) was applied to simulate the exchange-correlation potential [54]. van der Waals (vdW) dispersive interactions were included via the Tkatchenko–Scheffler (TS) scheme [55] to reliably predict the structural parameters in the $A_2BX_6$ perovskites due to contribution of vdW forces between the $[BX_6]^{2-}$ octahedra [56]. Calculations were performed including the influence of spin-orbit coupling (SOC) due to the presence of the heavy Sn element. We used the primitive unit cell of the $Rb_2SnBr_6$ in the $Fm\bar{3}m$ space group and convergence tests were applied using 330, 650 and 900 eV of plane-wave cutoff energies along with $(2 \times 2 \times 2)$ and $(10 \times 10 \times 10)$ Monkhorst-Pack k-point meshes [57]. The cutoff energy of 650 eV and k-point mesh of $(2 \times 2 \times 2)$ were confirmed to guarantee the convergence threshold criteria considered of $0.5 \times 10^{-5}$ eV/atom for total energy change, 0.01 eV/Å for total force per atom, 0.02 GPa for crystal pressure and maximum atomic displacement smaller than $0.5 \times 10^{-3}$ Å. For numerical optimization, the Broyden-Flether-Goldfarb-Shanno (BFGS) algorithm was applied [58]. For simulation of core electrons, the Norm-Conserving Pseudopotentials (NCPP) [59] were used with following valence configurations: $3d^{10}\,4s^2\,4p^5$ for Br atom, $4s^2\,4p^6\,5s^1$ for Rb atom and $4d^{10}\,5s^2\,5p^2$ for Sn atom. The electronic minimization parameters applied in the Self-Consistent Field (SCF) cycles were $0.5 \times 10^{-6}$ eV for total energy per atom and $0.1406 \times 10^{-6}$ eV for eigen-energy tolerance with three cycles for convergence tolerance window. Geometry optimization results agreed well with experimental values, displaying $a^{DFT} = 10.5295$ Å, with a deviation of ~ 0.13 Å (1.26 %) (see Fig. S1 of the Supplemental Material [60] for the structural parameters obtained for the other cutoff energies and k-point meshes mentioned in this section). After convergence was reached, the Kohn-Sham band structure and density of states (DOS) were calculated. Phonon properties were calculated using the Density Functional Perturbation Theory (DFPT) [61] within the linear response formalism with GGA(PBE)+TS level of theory upon the optimized unit cell with a q-point grid interpolation spacing of 0.05 $Å^{-1}$ and a convergence tolerance of $1\times10^{-5}$ eV/$Å^2$ on the interatomic force constants. To ensure smooth phonon dispersion curves, a fine interpolation with a point separation of 0.005 along the high-symmetry paths in the Brillouin zone was employed. For the electronic band structure, since pure functionals like GGA(PBE) tends to underestimate the energy bandgap [62], we have employed the Heyd – Scuseria – Enzerhof hybrid functional (HSE06) to improve the description of electronic structure and the bandgap prediction [63–68]. Apart of the energy bandgap value, similar trends were observed in the electronic band structures calculated with HSE06 and GGA(PBE)+TS+SOC, as shown in Fig. S2 [60]. We based the discussion in this work on the results obtained from the HSE06 calculations. Due to computational costs, SOC was not included in the phonon properties calculations or in the electronic band structure calculations using HSE06. The conductivity effective masses of electrons and holes were calculated from the harmonic average [69,70] of their respective effective masses from the valence band maximum (VBM) and conduction band minimum (CBM) along the $\Gamma \rightarrow X$, $\Gamma \rightarrow K$ and $\Gamma \rightarrow L$ directions of the high symmetry points as illustrated in Fig. S3 [60].

## III. RESULTS AND DISCUSSION

Fig. 1(a) shows the crystalline structure of the 0D perovskite $Rb_2SnBr_6$ at room conditions, which belongs to the $Fm\bar{3}m$ space group. This structure is derived from the 3D $RbSnBr_3$ perovskite [48,71] by withdrawing half of Sn atoms that occupy the $[SnBr_6]^{2-}$ octahedra centers, with $Rb^+$ ions embracing 12-fold coordination at sites among the $[SnBr_6]^{2-}$ units. The measured Sn-Br bond length of 2.5938(4) Å is constant along all octahedra, indicating a lack of bond distortions in the $[SnBr_6]^{2-}$ octahedron. Table SI of the Supplemental Material [60] summarizes the crystal data and parameters from SCXRD data refinement.

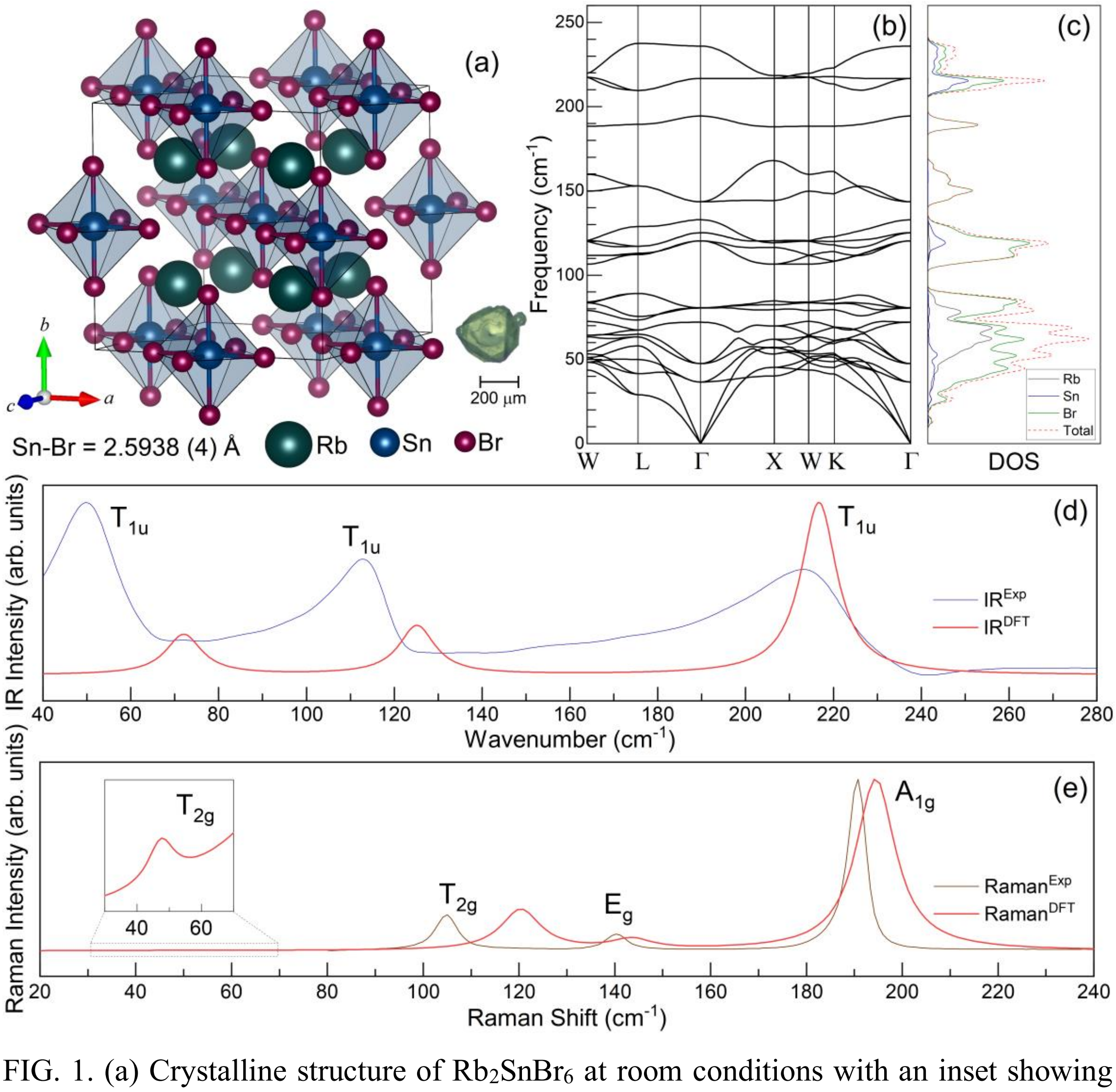

FIG. 1. (a) Crystalline structure of $Rb_2SnBr_6$ at room conditions with an inset showing one of the microcrystals synthesized in this work. (b) Phonon dispersion across the whole Brillouin zone and (c) phonon DOS of $Rb_2SnBr_6$ calculated at the GGA(PBE)+TS level

of theory. The observed and calculated optical modes are displayed in the normalized spectrum of (d) IR reflectance and (e) Raman spectroscopy.

The group theory with basis on the site occupation of $Fm\bar{3}m$ space group from $Rb_2SnBr_6$ predicts ten normal modes, which can be decomposed among the irreducible representation of the $m\bar{3}m$ factor group as: $\Gamma_M = A_{1g} \oplus E_g \oplus T_{2u} \oplus 2T_{2g} \oplus 4T_{1u} \oplus T_{1g}$. Among these modes, those with symmetries $T_{1u}$ and $T_{2u}$ are silent, one with symmetry $T_{1u}$ is acoustic, three with symmetry $T_{1u}$ are IR active, while the modes with symmetries $A_{1g}$, $E_g$ and $2T_{2g}$ are Raman active. Fig. 1(b) displays the calculated phonon dispersion by DFT along high-symmetry points $(\Gamma, L, K, W, X)$ in the Brillouin zone (see the Experimental and Computational Methods section for details about the first-principles calculations). The vibrational frequencies span from 0 to 250 $cm^{-1}$, with distinct separation of vibrational branches. The acoustic modes exhibit typical linear behavior near the Γ-point, with the longitudinal acoustic (LA) and transverse acoustic (TA) modes displaying moderate dispersion, reflecting the rigidity and isotropy of the crystal lattice [72,73]. The optical branches show significant splitting due to the pronounced mass contrast between Rb and Sn/Br and the macroscopic electric field inherent to the polar nature of $Rb_2SnBr_6$, a behavior observed in $CsBI_3$ (B = Pb, Sn) [74]. The absence of imaginary (soft) modes in the phonon dispersion indicates that the $Fm\bar{3}m$ crystal structure of $Rb_2SnBr_6$ is dynamically stable [75]. Fig. 1(c) highlights the phonon DOS contributions of Rb, Sn, and Br atoms to the vibrational modes. Low-frequency modes below 50 $cm^{-1}$ are primarily associated with the lattice vibrations dominated by the rattling motion of Rb atoms. Frequency modes in the 50–150 $cm^{-1}$ range are attributed to bending vibrations of the $[SnBr_6]^{2-}$ octahedra, with contributions from both Sn and Br atoms. LO and TO splitting are observed around this region, while frequency modes above 150 $cm^{-1}$ are dominated by stretching vibrations of the Sn–Br bonds, which reveals relatively low dispersion.

The normalized Far-IR reflectance spectrum of $Rb_2SnBr_6$ at room temperature, which is shown in Fig. 1(d), shows the three IR modes predicted from theory [76]: the highest wavenumber mode $\nu_{as}(T_{1u})$ corresponds to asymmetric stretching of Sn-Br bonds, the intermediate wavenumber vibration $\delta_{as}(T_{1u})$ represents the Br-Sn-Br asymmetric bending, and the lowest wavenumber mode $\nu^L(T_{1u})$ is related to a lattice mode with out-of-phase displacements of $[SnBr_6]^{2-}$ octahedra and $Rb^+$ cations. To obtain the frequencies of TO and LO modes, we used the Kramers-Kronig relations [77] on the reflectance spectra to obtain the real $Re(\varepsilon)$ and imaginary $Img(\varepsilon)$ parts of the dielectric function $\varepsilon(\omega)$. The TO modes correspond to peaks in $Img(\varepsilon)$, whereas maxima in the imaginary part of the inverse of the dielectric function $-Img(1/\varepsilon)$ represent the LO modes. Both $Img(\varepsilon)$ and $-Img(1/\varepsilon)$ are shown in Fig. S4 [60]. The obtained wavenumber values of the LO and TO modes are listed in Table SII [60]. To obtain the high-frequency dielectric constant $\varepsilon_\infty$, we extrapolated $Re(\varepsilon)$ at high frequencies and estimated $\varepsilon_\infty$ = 3.71. The static dielectric constant $\varepsilon_s$ was calculated through the generalized Lyddane-Sachs-Teller relation [77]:

$$\frac{\varepsilon_s}{\varepsilon_\infty} = \prod_i \left(\frac{\omega_{LO,i}}{\omega_{TO,i}}\right)^2 \tag{1}$$

where $\omega_{TO}$ and $\omega_{LO}$ are the TO and LO frequencies, respectively. Using the experimental values of the TO and LO modes in Eq. (1), $\varepsilon_s$ was evaluated as 7.57. These parameters are in accordance with previously reported theoretical values for isomorphous compounds $Cs_2BX_4$ (B = Sn, Ti; X = Cl, Br, I) [56].

Fig. 1(e) presents the room-temperature Raman spectrum of $Rb_2SnBr_6$ under 457 nm excitation wavelength (2.71 eV), which consists mainly of three intense peaks linked to modes of $[SnBr_6]^{2-}$ vibrations that can be described, from lower to higher Raman shift, as follow [76,78]: $\delta_{as}(T_{2g}) = 105\ cm^{-1}$ asymmetric bending of Br-Sn-Br bonds, $\nu_{as}(E_g) = 140\ cm^{-1}$ asymmetric Sn-Br stretching and $\nu_s(A_{1g}) = 191\ cm^{-1}$ symmetric stretching of Sn-Br. A fourth weak intensity peak at a higher Raman shift was assigned as the second-order mode of $\nu_s(A_{1g})$. The other $\nu^L(T_{2g})$ Raman active mode, corresponding to vibrations of $Rb^+$ cations within fixed $[SnBr_6]^{2-}$ lattice framework, is supposed to be at Raman shift values lower than 80 $cm^{-1}$ (out of measured range). This mode aligns with the low-intensity calculated Raman DFT phonon around 50 $cm^{-1}$. These observations confirm strong agreement between experimental and DFT-calculated Raman and IR modes, with all frequencies aligning with the calculated phonon dispersion and phonon DOS. Summarized values are provided in Table SII [60].

We performed low-temperature Raman spectroscopy to investigate possible structural phase transitions (SPT) or other anomalies in $Rb_2SnBr_6$. However, no substantial changes consistent with SPT or other anomalies were observed aside from expected thermal lattice anharmonicity accompanied by a narrowing of Raman peaks upon cooling, as shown in Fig. S5 [60]. The changes in Raman shift modes from room-temperature to low-temperatures were notably small, with the highest difference less than 5 $cm^{-1}$ (0.6 meV) for the $\nu_s(A_{1g})$, showing that the cubic structure and bonding strength persist at lower temperatures [33]. This is expected because the vacancy-ordered structure maintains the configuration of isolated octahedra, preventing phase transitions induced by tilting corner-sharing octahedra, commonly occurring in Pb-based halide perovskites during lattice contraction [79,80].

Fig. 2(a) presents the calculated electronic band structure of $Rb_2SnBr_6$ by DFT (see the Experimental and Computational Methods section for details about the DFT-based theoretical calculations). These 0D MHPs exhibit a direct electronic transition at the Γ-point (3.09 eV). From the DOS of the electronic band structure shown in Fig. 2(b), the valence band maximum (VBM) and conduction band minimum (CBM) are derived from the splitting of $[SnBr_6]^{2-}$ octahedra molecular orbital, with a nonbonding Br *p*-state at the VBM, and single bonding Sn-Br (*s*-*p*) state at the CBM, similar to $Cs_2SnBr_6$ [36,56]. The CBM has a gradient band edge from the mixing and delocalization of Sn $s$ and Br $p$ states, while the VBM presents less dispersive features from the fully oxidized $Sn^{4+}$, unlike conventional 3D tin halide with partially oxidized $Sn^{2+}$ and filled valence subshell [56], indicating the presence of light electrons and heavy holes as can be seen from the conductivity effective masses $\overline{m_e}$ and $\overline{m_h}$ of electron and hole, respectively, with values of $\overline{m_e} = 0.445m_0$ and $\overline{m_h} = 1.12m_0$ where $m_0$ is the electron rest mass. To experimentally determine the optical bandgap of $Rb_2SnBr_6$, we conducted UV-Vis DRS in polycrystalline samples at room temperature with the obtained reflectance spectrum $R$ converted to absorbance units using the Kubelka-Monk function $F(R)$ as shown in Fig. 2(c). We can see that the optical absorption edge is around 450 nm. We estimated the energy bandgap by extrapolating the linear direct-Tauc plot [81], given by $[F(R)E]^2 \propto (E - E_g)$ where $E$ is the energy and $E_g$ stands for the direct energy bandgap, obtaining $E_g = 2.960\ (1)\ eV$. This experimental bandgap aligns remarkably well with the theoretical bandgap obtained through DFT calculations and it is in good agreement with other isomorphous perovskites [82–84].

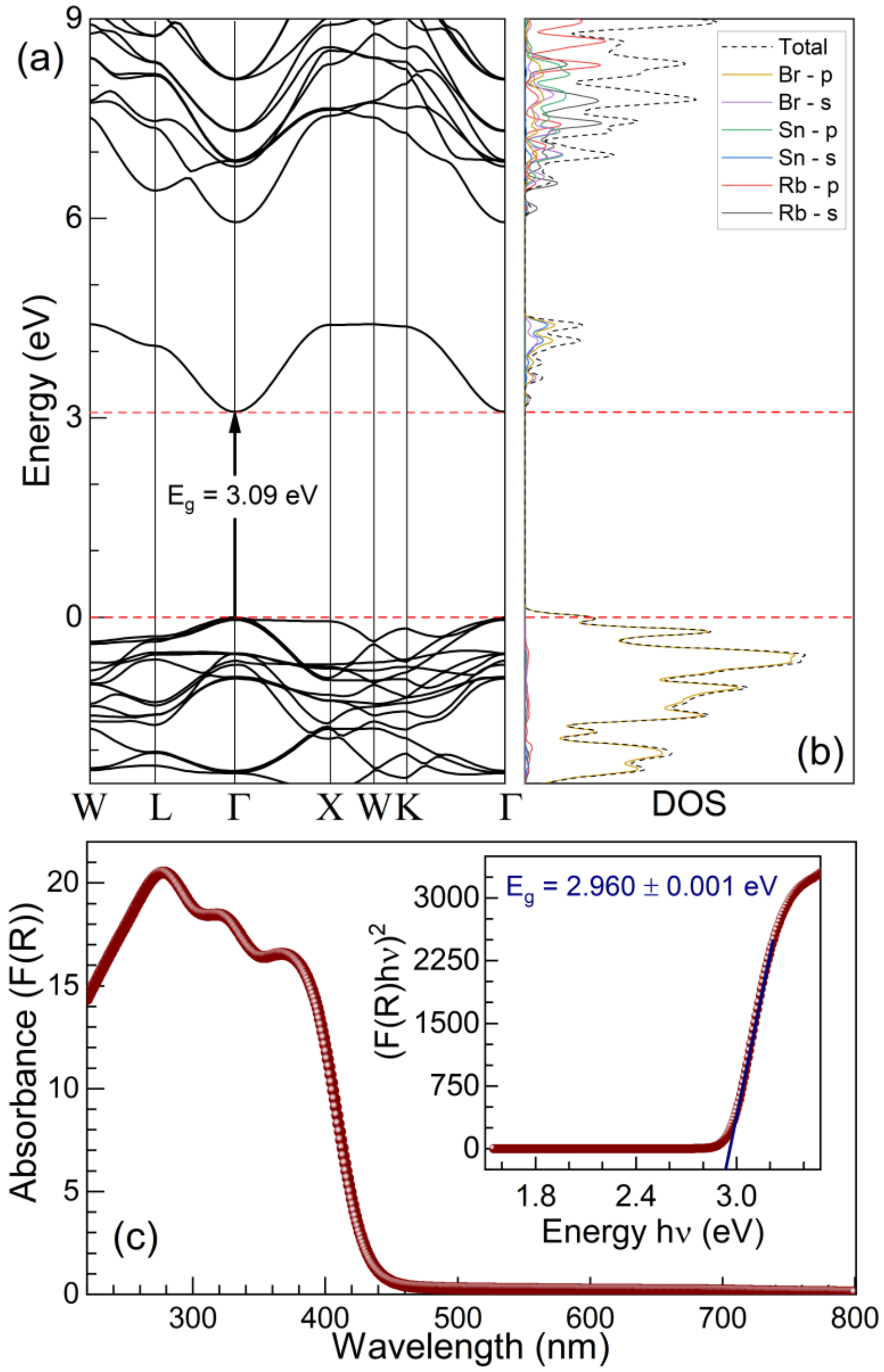

FIG. 2. (a) Electronic band structure and (b) electronic DOS of $Rb_2SnBr_6$ calculated at the HSE06 level of theory. (c) Normalized (measured) absorbance spectrum of $Rb_2SnBr_6$ with the direct-Tauc plot in the inset. The calculated bandgap (3.09 eV) is consistent with the experimental optical bandgap (2.960 eV).

The PL spectrum of $Rb_2SnBr_6$ at 10 K under UV excitation (3.06 eV) showed in Fig. 3(a) revealed a broad emission around 1.93 eV (642 nm) and a full width at half maximum (FWHM) of 202 meV, which can be modeled by a Gaussian peak. Such broadband and large Stokes shift energy is usually found in low-dimensional halide perovskites [82,85,86]. They are usually attributed to a wide distribution of states inside the bandgap that behave as traps to charge carriers where below-gap photons can be emitted [87,88] or to STE luminescence where strong EPC plays a crucial role creating polarons from lattice distortions that bind charge carries of an exciton to the lattice (self-trapping) at in-gap energy levels with subsequent phonon-assisted radiative emission from exciton recombination [87,89,90].

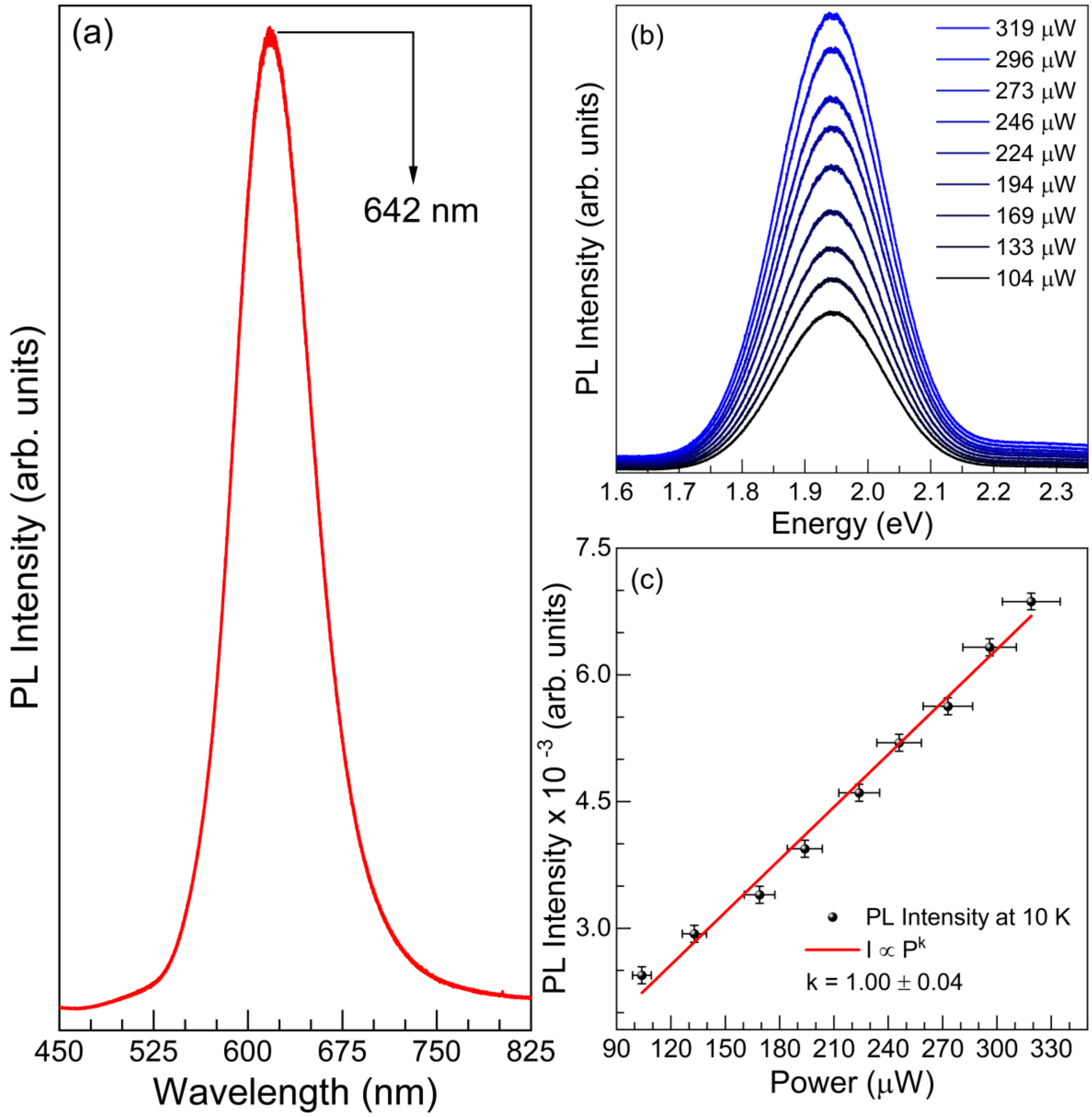

FIG. 3. (a) PL spectrum at 10 K of $Rb_2SnBr_6$. (b) PL emission dependence of $Rb_2SnBr_6$ PL at 10 K on excitation power. (c) A plot of PL intensity vs power was obtained from (b).

To further verify the origin of the $Rb_2SnBr_6$ emission band, we performed the PL intensity dependence on excitation power at 10 K, as illustrated in Fig. 3(b). An increase in intensity and a small redshift of 1 meV on the PL peak was observed while the excitation power increased from 104 μW to 319 μW. The general relation of PL intensity $I$ and excitation power $P$ is described as $I \propto P^k$, where the exponent $k$ lies on the ranges $0 < k < 1$ for free-to-bound transitions or donor-acceptor pairs (DAP) recombination and $1 < k < 2$ for excitonic emission [91]. Excitation energies close to the bandgap energy tend to result in a linear relation between $I$ and $P$ for the excitonic recombination process that agrees with our data, as seen in Fig. 3(c), which shows the plot of $I$ vs $P$, yielding $k = 1.00 \pm 0.04$. The redshift behavior contrasts with the characteristic blueshift of DAP recombination, which excludes this possibility of emission mechanism [87].

Additionally, free-to-bound transitions are identified as acceptor or donor levels that bind carriers and become ionized at higher temperatures, leading to emission bands from free carriers [32,92]. That is not observed on the temperature-dependent PL spectra in Fig. 4(a) once the peak energy and intensity monotonically decrease from 10 K until its quenching around 180 K with no other radiative emission bands detected. Thus, free-to-bound recombination should not be the mechanism of luminescence found in our results. In this way, we suggest that exciton recombination is responsible for PL emission of $Rb_2SnBr_6$.

The PL emission color, under UV excitation upon increasing temperature from 10 K, changed from orange to light-yellow until 180 K, where no PL emission could be detected (see the inset of Fig. 4(a) that shows the dependence of PL peak energy with temperature and the CIE plot in Fig. S6 [60] with chromaticity coordinates in Table SIII [60] of the Supplemental Material). Moreover, the PL thermal quenching behavior was modeled considering the integrated PL intensity $I_{PL}(T)$ as a function of the temperature $T$ using the Arrhenius equation [93,94] given by

$$I_{PL}(T) = \frac{I_0}{1 + Ae^{\frac{-E_a}{k_B T}}} \quad (2)$$

where $k_B$ is the Boltzmann constant, $I_0$ stands for zero-temperature integrated intensity, $A$ corresponds to the density of radiative recombination centers, and $E_a$ relates to the activation energy for thermal quenching. Eq. (2) fully agreed with our data, as it is shown in Fig. 4(b), estimating an activation energy $E_a = 16.7 \pm 0.7$ meV, close to the quenching temperature of 180 K ($\approx$ 16 meV). The absence of emission beyond 180 K suggests that the calculated activation energy is associated with the nonradiative recombination process.

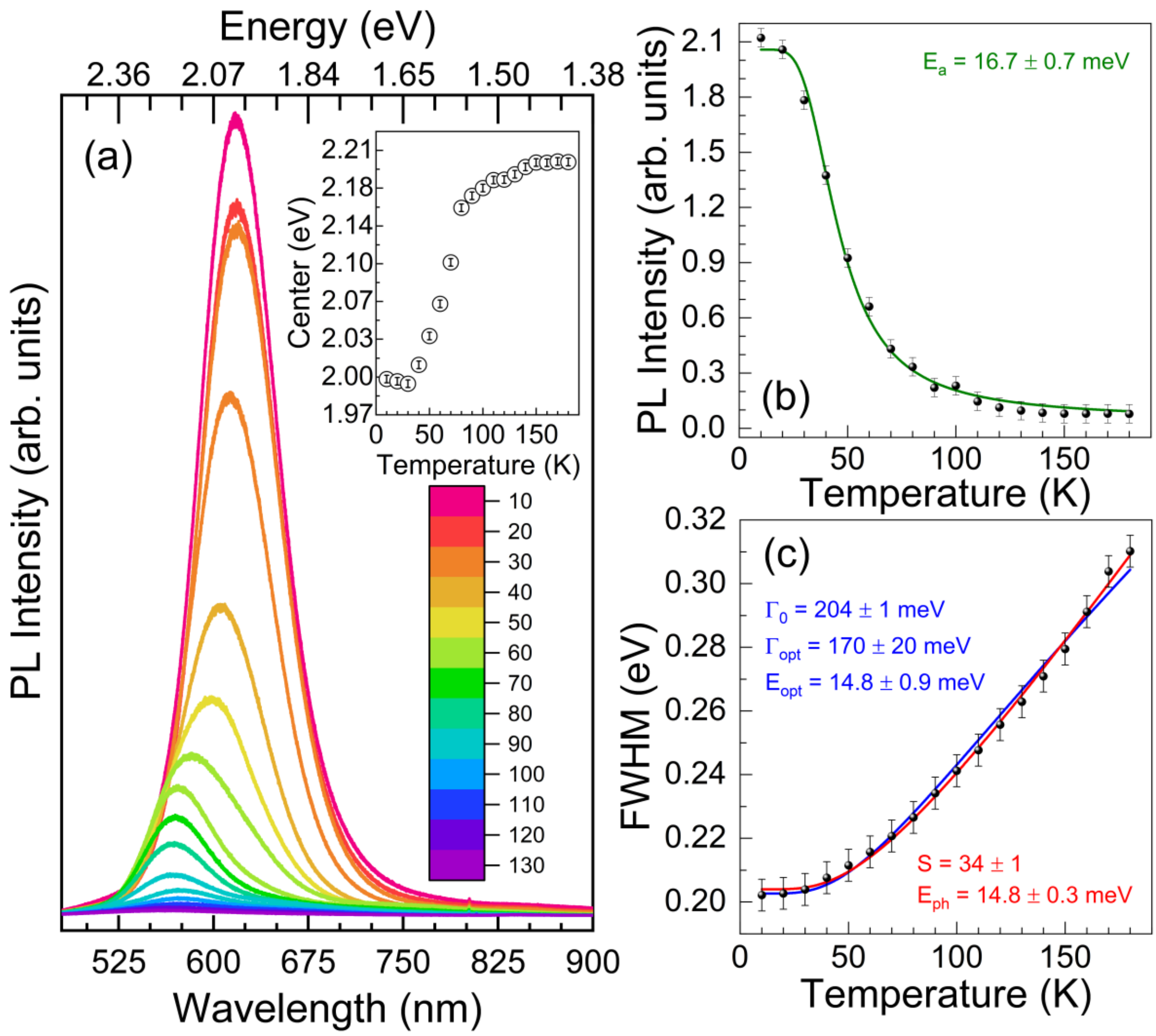

FIG. 4. (a) Temperature-dependent PL spectra of $Rb_2SnBr_6$. The inset shows the temperature dependence of peak energy. (b) Integrated PL intensity *vs* temperature fitted with Eq. (2). (c) PL FWHM vs temperature analyzed by Eq. (3) (solid blue line) and Eq. (4) (solid red line).

To evaluate the EPC strength involved in the luminescence phenomena of $Rb_2SnBr_6$, we investigated the temperature dependence of PL FWHM presented in Fig. 4(c) based on two perspectives. The first model used to examine the PL linewidth was elaborated by Rudin *et al*. [95] and concerns different contributions to PL broadening on temperature, which can be represented as the sum of functional dependencies into FWHM:

$$FWHM(T) = \Gamma_0 + \Gamma_{ac}T + \frac{\Gamma_{LO}}{e^{\frac{E_{LO}}{k_BT}} - 1} + \Gamma_{imp}e^{\frac{-E_{imp}}{k_BT}} \qquad (3)$$

In the Eq. (3), the first term $\Gamma_0$ is the inhomogeneous broadening arising from scattering due to lattice disorder [24]; the second term $\Gamma_{ac}$ the term represents the homogeneous broadening from the coupling between excitons and acoustic phonons through deformation potential scattering where lattice distortions modify the electronic band structure [96]; the third term $\Gamma_{LO}$ stands for the homogenous broadening contribution of LO phonons of energy $E_{LO}$ via long-range Fröhlich interactions, which is proportional to the occupation number of phonons given by Bose-Einstein distribution $1/(e^{\frac{E_{LO}}{Kt}} - 1)$ and emerges from the interplay between the macroscopic electric field induced by LO vibrations and exciton's charge carries [33]; the last term $\Gamma_{imp}$ accounts for inhomogeneous broadening due scattering from ionized impurities with mean binding energy $E_{imp}$ [97]. In general, the homogeneous and inhomogeneous broadening lineshapes are best described by Lorentzian and Gaussian functions, respectively, and their convolution yields the overall PL lineshape [98,99]. The model described by Eq. (3), when applied to our data, provides insights into the dominant contributions to the temperature dependence of the PL linewidth.

Based on our previous analysis about the PL intensity dependence on excitation power that excluded DAP or free-to-bound recombination process and the linear increase of FWHM experimental data as it approximates the quenching temperature, which does not agree with the asymptotic behavior expected from the exponential term associated with ionized impurities, we assumed $\Gamma_{imp} = 0$. At the low-temperature range, the optical phonon contributions from the Bose-Einstein distribution function give a gradient of zero in the regime of $E_{LO} < k_B T$. However, on a typically observable range of temperature, as achieved in this work, acoustic phonons have significantly lower energies than $k_B T$ and a non-zero gradient of $\Gamma_{ac}$ is predicted as $T$ approaches zero [24]. Observing the FWHM dependence on temperature, we see that the gradient tends to vanish for the lowest temperatures, indicating negligible contribution from acoustic modes ($\Gamma_{ac} \approx 0$) which was further verified with the fit convergence when $\Gamma_{ac} \rightarrow 0$. Therefore, we concluded that Fröhlich interactions through LO phonons play a major role in PL broadening temperature dependence.

Thus, fitting the PL FWHM with Eq. (3) considering only the coupling between excitons and LO phonons from the Fröhlich mechanism and the temperature-independent inhomogeneous broadening, we obtain $\Gamma_0 = 204 \pm 1$ meV, $\Gamma_{LO} = 170 \pm 20$ meV and $E_{LO} = 14.8 \pm 0.9$ meV. The energy $E_{LO}$ is closely related with $\delta_{as}(T_{1u})$ LO phonon mode at 120 cm$^{-1}$ (≈ 15 meV) obtained by IR analysis, which suggests that long-range Fröhlich interactions mediated by LO phonons are the dominant mechanisms of EPC that impact PL broadening, that is commonly expected in polar semiconductors as is the case for $Rb_2SnBr_6$ [10,100]. In the present case, the $\Gamma_{LO}$ parameter measures the strength of electron-phonon coupling that is notably higher than in 3D perovskites [24,101], shedding light on the strong strength of EPC in low-dimensional perovskite compounds.

The second model used to evaluate the EPC strength was developed by Toyozawa [102,103] employing a configuration coordinate model where the mean number of phonons with effective energy $E_{ph}$ emitted after exciton recombination is described by the Huang-Rhys factor $S$ [29] in the form of

$$FWHM(T) = 2.36\sqrt{S}E_{ph}\sqrt{\coth\left(\frac{E_{ph}}{2k_BT}\right)} \quad (4)$$

Here, the value of $S$ determines the regime of EPC. Generally, $S \gg 1$ implies a strong coupling regime [87]. In Toyozawa's theory, the case of strong electron/exciton-phonon coupling would lead to a Gaussian shape of PL linewidth rather than a Lorentzian one, that happens in the weak couple regime [89], which is consistent with our findings. The temperature-dependent FWHM fitted with Eq. (4) yielded $E_{ph} = 14.8 \pm 0.3$ meV and $S = 34 \pm 1$. The effective phonon energy matches with the LO phonon energy $E_{LO}$ of the $\delta_{as}(T_{1u})$ mode previously found with Rudin's model, supporting that the EPC of $Rb_2SnBr_6$ arises from Fröhlich interactions with LO phonons. The strong EPC revealed by the Huang-Rhys factor agrees with the high value obtained from $\Gamma_{LO}$ and is compatible with the assumption of exciton self-trapping underlying the below-gap broad luminescence from STE as observed in many halide perovskites available in the literature [30,32,82,104].

In the Fröhlich theory, the polaron formation arises from the interactions between carriers and LO phonons, and the coupling strength of this carrier-vibration relation is determined by the Fröhlich parameter $\alpha$ that is defined as

$$\alpha = \frac{e^2}{4\pi\varepsilon_0\hbar}\left(\frac{1}{\varepsilon_\infty} - \frac{1}{\varepsilon_s}\right)\sqrt{\frac{\bar{m}}{2\hbar\omega}} \tag{5}$$

where $e$ is the electron charge, $\varepsilon_0$ is the vacuum permittivity, ħ is the reduced Planck constant, $\omega$ is the LO phonon frequency, and $\bar{m}$ is the conductivity effective mass of the electron or hole. Using the first-principles calculated effective masses as well as the experimental values for $\hbar\omega$ (LO phonon energy), $\varepsilon_\infty$ and $\varepsilon_s$, Eq. (5) yielded Fröhlich parameters of 2.78 and 4.41 for the electron and hole, respectively. Although the calculated $\alpha$ for the electrons is the same range of $1 - 3$ [10,27,100] for conventional 3D halide perovskites (for both electrons and holes), we note that the calculated hole-phonon coupling is well above this range as a direct consequence of the heavy hole mass compared to the electron found through DFT calculations. Upon hole interactions with surrounding lattice, its mass is further increased by the hole-polaron mass $\bar{m}_{hp} \approx \left(1 + \frac{\alpha}{6} + \frac{\alpha^2}{40}\right)\bar{m}_h = 2.49m_0$ that aid hole localization and consequently exciton trapping by its positive carrier [104]. Thus, we proposed that STE formation in $Rb_2SnBr_6$ is mediated by hole localization upon Fröhlich interactions with the lattice vibration. This kind of exciton formation has been observed in organic and molecular crystals [105], and total inorganic salts such as AgCl [87,106].

The above results about the strong EPC of low-dimensional $Rb_2SnBr_6$ compared with 3D halide perovskites can be interpreted in terms of the [$BX_6$] octahedra that are most responsible for the electronic and optical properties of many perovskite materials [2,31,107]: while in 3D perovskites the octahedra are surrounded by a corner-sharing framework, isolated [$BX_6$] octahedra configuration of 0D perovskites possess more degrees of freedom and hence are easily distorted upon interaction between exciton's charge carriers (in the present case, the hole) and octahedra vibrations creating polarons that are dynamically related to LO phonons due the polar nature of the perovskites [21,108]. Thus, the enhancement of EPC promotes exciton trapping and STE formation, which leads to a broad luminescence with large Stokes shift energy, as shown by our findings and data analysis.

## IV. CONCLUSIONS

We investigated the electron-phonon coupling in 0D $Rb_2SnBr_6$ microcrystals, analyzing their structural, vibrational, and optical properties through several experimental techniques and theoretical calculations. Specifically, single crystal X ray diffraction revealed a cubic symmetry in $Fm\bar{3}m$ space group with a direct energy bandgap $E_g = 2.960(1)\ eV$ in the UV-Vis range obtained by diffuse reflection spectroscopy. Infrared and Raman spectroscopies showed the main vibrational modes associated with bending and stretching of $[SnBr_6]^{2-}$ octahedra, and low-temperature Raman analyses provided no evidence of any SPT down to 10 K. DFT results were very consistent with the vibrational and optical properties found at room temperature. Low-temperature photoluminescence (PL) spectrum indicated a broad emission with large Stokes shift energy consistent with excitonic self-trapping, as the power and temperature-dependent PL pointed out. The EPC was evaluated upon the PL linewidth broadening temperature dependence based on two perspectives where charge carriers interplay with Br-Sn-Br asymmetric bending of LO phonons mediated by Fröhlich interactions were the main mechanism responsible for the PL broadening through phonon-assisted radiative emission. Strong EPC was elucidated by the Huang-Rhys factor $S = 34 \pm 1$ coherent with STE emission commonly found in low-dimensional perovskites. The Fröhlich parameter $\alpha$ found through experimental data, and theoretical calculations pointed out the main contributions of hole localization by large hole-polaron mass in STE formation. Our work provides a fundamental understanding of the correspondence between vibrational and optoelectronic properties of $Rb_2SnBr_6$, guiding its applications such as light-harvesting devices.

## AUTHOR CONTRIBUTIONS

C. C. S. S. and J. S. R. H. conceptualized the project. C. C. S. S. wrote the original draft. Experimental work was conducted by C. C. S. S, J. S. R. H, V. S. N., and M. A. P. G, performing the single crystal x-ray diffraction, diffuse reflectance spectroscopy, Raman spectroscopy, IR spectroscopy, and photoluminescence. B. P. S performed the first-principles calculations. The data analysis was conducted by C. C. S. S, and J. S. R. H. Supervision, resources, and funding acquisition were acquired by C. W. A. P. and A. P. A. All authors participated in reviewing and interpreting the results.

## ACKNOWLEDGEMENTS

This investigation was financed in part by Fundação Cearense de Apoio ao Desenvolvimento Científico e Tecnológico (UNI-0210-00047.01.00/23, 07548003/2023), the Conselho Nacional de Pesquisa e Desenvolvimento do Brasil (CNPq – Projetos 407956/2022-0, 407954/2022-8, 406322/2022-8, 310353/2023-8), and Coordenação de Aperfeiçoamento de Pessoal de Nível Superior do Brasil (CAPES) - Finance Code 001 (CAPES – Projeto 88887.669776/2022-00). The authors would like to thank Prof. Tiago Pinheiro Braga from the Physical–Chemistry Department of the Federal University of Rio Grande do Norte and the NPAD/UFRN for the computational resources provided. Additional calculations were conducted using the resources of the Centro Nacional de Processamento de Alto Desempenho in São Paulo (CENAPAD-SP), Project 823.

## DATA AVAILABILITY

The data that support the findings of this article are not publicly available upon publication because it is not technically feasible and/or the cost of preparing, depositing, and hosting the data would be prohibitive within the terms of this research project. The data are available from the authors upon reasonable request.